\documentclass[final,5p,times,twocolumn]{elsarticle}
\usepackage{graphicx}
\usepackage{enumerate}

\usepackage{subfigure}
\usepackage{amsmath}
\usepackage{amssymb}
\usepackage{amsfonts}
\usepackage{float}
\usepackage{caption}
\usepackage{color}
\usepackage{array}

\newcolumntype{?}{!{\vrule width 1pt}}
\makeatletter
\def\hlinewd#1{%
  \noalign{\ifnum0=`}\fi\hrule \@height #1 \futurelet
   \reserved@a\@xhline}
\makeatother

\providecommand{\apj}[0]{ApJ}
\providecommand{\apjl}[0]{ApJ Lett.}

\providecommand{\prd}{PRD}

\def\beq{\begin{equation}}
\def\eeq{\end{equation}}

\begin{document}

\begin{frontmatter}
\title{Detector Optimization Figures-of-merit for \\ IceCube's High-energy Extension}

\author{I. Bartos}
\ead{ibartos@phys.columbia.edu}
\address{Department of Physics, Columbia University, New York, NY 10027, USA}

\begin{abstract}
With the design and development of next-generation high-energy neutrino detectors, it is important to compare different detector designs to optimize detection probability and science reach. These comparisons are nevertheless difficult due to large uncertainties in current neutrino source model parameters. We examine the role of the most important characteristics of high-energy neutrino searches in the probability of discovering different sources types. We derive scaling relations for each considered source and search scenario, which can be used to compare different detector designs with respect to their utility in discovering different source populations. The recovered scaling relations are independent of source strengths, providing a model-independent comparison.
\end{abstract}





\end{frontmatter}

\section{Introduction}

IceCube's recent discovery of cosmic high-energy neutrinos \cite{2013Sci...342E...1I,2014PhRvL.113j1101A,2015PhRvL.115h1102A,2015ApJ...809...98A} represents an important step in studying high-energy astrophysical processes with neutrinos. Nevertheless, the origin of the observed neutrinos is currently unknown. Multimessenger searches have not recovered electromagnetic \cite{2015ApJ...805L...5A,2015ApJ...811...52A} or gravitational wave \cite{2014PhRvD..90j2002A} counterparts, and the neutrinos' directional distribution so far is consistent with diffuse emission \cite{2015APh....66...39A}. The observed neutrino energy distribution is also so far consistent with multiple possible source types \cite{2008ApJ...689L.105M,2009ApJ...707..370K,2015ApJ...805...95C,2015ApJ...806...24S,2015arXiv150900983B}.

Some high-energy neutrino observatories are planned to be substantially expanded in the near future. IceCube's planned upgrade, IceCube-Gen2 \cite{2014arXiv1412.5106I}, aims to instrument 10\,km$^3$ volume, essentially increasing sensitivity by a factor of 10. The KM3NeT detector is planned to be constructed in the Mediterranean with comparable volume \cite{2010NIMPA.623..445D}. There are also plans to expand the Baikal neutrino detector at lake Baikal to km$^3$ volume \cite{2011NIMPA.628..115A}. 

The specifics of how a neutrino observatory is built or expanded given a fixed cost depend on the relative importance of several factors. For example, a part of the detector can be dedicated to help reject the atmospheric background. This part then, however, is not used in direct detection, therefore reducing the overall sensitivity. Such tradeoffs can be made by aiming to optimize the science reach of the detector. 

The goal of this paper is to assess how detector characteristics affect the science reach of a high-energy neutrino detector. Given the limited resources for construction, the optimization of the detector's ability to probe the high-energy universe is essential. We will focus on the discovery of the astrophysical sources of origin of cosmic neutrinos. For numerical values, we will consider IceCube's Gen2 extension, but the results are generally valid for other detectors as well.

While there are numerous critical aspects of detector construction, here we approach this question from the perspective of some of the fundamental detector characteristics, such as (i) sensitivity, (ii) directional uncertainty, and (iii) background rejection ability. While these quantities can vary with source direction, time, etc, here we consider them as characteristic overall values and describe the detector's capability with them. While local variations of these parameters may affect search sensitivity, they are unlikely to change the general relations that we derive below. There are also other important parameters, such as the reconstructed energy uncertainty, which can play an important role in, e.g., spectral analysis. The search method we consider in the following takes advantage of directional and temporal correlation, and therefore energy reconstruction will not be of primary importance.
Our aim is to derive scaling relations between the above three characteristic parameters that determine the detector's sensitivity for different source types. Such a relation can then help establish a detector design that maximizes the chance of discovery within the available resources. In other words, it is likely that the design process requires compromises between achieving greater sensitivity, directional uncertainty or background rejection capability. This work tries to help make the best compromise.

There are different source types for which detector design optimization may be different. We will consider
\begin{enumerate}
\item continuous point sources
\item transient sources
\item an extended galactic source
\end{enumerate}
Further, we will examine searches for TeV neutrinos for which a large number of events are detected with IceCube, as well as ultra-high energy neutrinos with a significantly lower number of detections and better signal-to-noise ratio.

We are interested in the connection between the probability of discovering a given source type, and the following parameters:
\begin{enumerate}[$\circ$]
\item \boldmath{$N_{\rm ast}$}: number of astrophysical neutrinos detected during the considered observation period.
\item \boldmath{$\psi$}: angular uncertainty of the reconstructed neutrino direction.
\item \boldmath{$N_{\rm atm}$}: number of atmospherical background events detected during the considered observation period.
\end{enumerate}
We will derive a scaling relation between these three quantities for each of the considered source type.

The paper is structured as follows. Section \ref{section:searchstrategy} introduces a search strategy that will be assumed in deriving detection probabilities. Sections \ref{section:contpoint}, \ref{section:transients} and \ref{section:galactic} describe results for continuous point sources, transients and extended galactic sources, respectively. Section \ref{section:summary} summarizes and discusses the results.

\section{Search strategy}
\label{section:searchstrategy}

There are a large number of possible search strategies for high-energy neutrinos for a variety of source types. Here we adopt a simple strategy that captures the important features of efficient multimessenger searches, while allowing for the analytical evaluation of the sensitivity.

Since we are interested in discovering the sources of origin, we consider a known astrophysical population of potential sources. Accordingly, the search relies on a catalog of potential neutrino sources observed with electromagnetic telescopes/detectors, with precisely known directions and distances, and if applicable, times and durations. For simplicity, we assume that the sources are uniformly distributed in the universe.

The strategy is the following. We count directional coincidences between the detected events and the source catalog. The significance of the search is determined based on the number of observed coincidences compared to the number expected assuming no correlation between neutrino and catalog directions.
Since a too large number of source in the catalog render the search not sufficiently sensitive, we exclude the farthest sources in the catalog such that the expected p-value from background-only neutrino events equals the p-value $p_{\rm disc}$ determined by the required false alarm rate. This ensures that, having a neutrino that originates from one of the sources within the (reduced) catalog will be a discovery. Note that the downside of the reduced catalog is that the detected astrophysical neutrinos that originate from the sources that are excluded from the reduced catalog are now part of the background.

We quantify the sensitivity of the above search strategy by the probability $P(p_{\rm disc})$ that neutrinos from a source population are discovered with p-value $\leq p_{\rm disc}$. This probability will also depend on the detector parameters.

We note that, while this search is relatively simple and does not take into account all information available in the source catalog, we find that its scaling is similar to that of a more complete search strategy, with comparable sensitivity. The reason is that at larger distances, sources are more uniformly distributed, therefore directional correlation gradually loses significance.

The derivation of the scaling is different for different scenarios. In the following, we organize the discussion along source types and neutrino energy ranges of interest, pointing out where the derivations overlap.

\section{Continuous point sources}
\label{section:contpoint}

Continuous point sources are expected to continuously emit neutrinos, from within an angular size much smaller than the angular uncertainty of reconstructed neutrino directions. This class includes some of the most promising neutrino source candidates, such as starburst galaxies, active galactic nuclei (AGN), and quasars. We will discuss extended galactic sources in a separate section.

In the following we will separately consider the cases of ultra-high-energy ($\gtrsim100$\,TeV) neutrino and TeV neutrino searches. These two cases have different signal and background detection rates that require separate treatment. Since we are interested in directional coincidence, for both cases we will focus on track events that have much better reconstructed directions than cascade events.

\subsection{Ultra-high-energy search for continuous point sources}
\label{section:UHEcontinuous}

\subsubsection{Number of detected signal and background neutrinos}

It is useful to introduce $N_{\rm s}$ and $N_{\rm b}$, the numbers of detected signal and background neutrinos, respectively. Signal neutrinos will be defined as those that originate from one of the sources in the reduced catalog. All other detected events, including atmospheric events as well as astrophysical neutrinos originating from sources not in the catalog, will be considered background, and their number will be denoted with $N_{\rm b}$. Since the search is limited to sources within $d_{\rm th}$, only a fraction
\begin{equation}
f_{\rm d} \propto d_{\rm th}
\label{eq:fd}
\end{equation}
of astrophysical neutrinos will be included in $N_{\rm s}$. This scaling comes from the fact that the neutrino contribution of a volume shell within $[d_{\rm th},d_{\rm th}+\epsilon]$ is independent of $d_{\rm th}$, where $\epsilon$ is a small distance. We therefore have
\begin{eqnarray}
N_{\rm s} &=& f_{\rm d} N_{\rm ast}  \label{eq:Ns} \\
N_{\rm b} &=& N_{\rm atm} + (1 - f_{\rm d}) N_{\rm ast} \\
N_{\rm total} &=& N_{\rm s} + N_{\rm b}
\end{eqnarray}

\subsubsection{Discovery probability}

We take advantage of the fact that ultra-high-energy searches have relatively few detections by optimizing for the detection of one signal neutrino. Since we focus on only the nearby sources (say within a few hundred Mpc) within which directional correlation can be significant, the fraction of astrophysical neutrinos originating from these nearby sources is small. Even with a $\sim10$ times increased detection rate, the number of astrophysical neutrinos originating from a source within $\sim 100$\,Mpc from a uniform source population over $\sim1$\,yr observation is $N_{\rm s} \ll 1$.

In the limit of $N_{\rm s} \ll 1$ the probability of discovery is
\begin{equation}
P(p_{\rm disc}) \approx 1 - \mbox{Pois}(0,N_{\rm s}) \approx N_{\rm s},
\label{eq:Pp}
\end{equation}
where $\mbox{Pois}(k,\lambda)$ is the Poisson distribution for $k$ with parameter $\lambda$.

\subsubsection{Threshold distance}

With $\psi$ directional uncertainty for the neutrino events and precise (i.e. $\ll\psi$) directional uncertainty for the source catalog, the chance of random directional coincidence between a neutrino and a source is $\propto \psi^2$. Let $\rho_{\rm source}$ be the number density of the astrophysical sources in the catalog, and $d_{\rm th}$ a threshold distance such that we only include sources in the catalog if they are within $d_{\rm th}$ to the observer. Let $N_{\rm total}$ be the total number of detected events, including astrophysical neutrinos and atmospheric background events. With these parameters, the probability of false discovery, i.e. the probability of a chance overlap between any of the background neutrinos and sources is
\begin{equation}
p \approx k \psi^2 \rho_{\rm source} d_{\rm th}^3 N_{\rm total}
\label{eq:p}
\end{equation}
where $k$ is an appropriate numerical constant. Note that this approximation is accurate for $p\ll 1$.

If we want a single neutrino from the source population will be discovery, we can choose $d_{\rm th}$ such that $p$ in Eq. \ref{eq:p} is equal to the discovery threshold $p_{\rm disc}$. The required distance threshold that will render a single signal neutrino a discovery is therefore
\begin{equation}
d_{\rm th} = p_{\rm disc}^{1/3} k^{-1/3} \psi^{-2/3} \rho_{\rm source}^{-1/3} N_{\rm total}^{-1/3}
\label{eq:dth}
\end{equation}

\subsubsection{Scaling relation}

We now combine Eqs. \ref{eq:fd}, \ref{eq:Ns}, \ref{eq:Pp} and \ref{eq:dth} to determine the scaling relation between $P(p_{\rm disc})$ and the detector parameters. We find
\begin{equation}
P(p_{\rm disc})|_{\rm continuous,UHE} \propto N_{\rm ast} \psi^{-2/3} N_{\rm total}^{-1/3}.
\end{equation}
This is the scaling relation we can use to compare detector designs with respect to the probability of discovery of continuous sources with searches for ultra-high-energy neutrinos. For example, if design A has twice the effective area but the angular uncertainty is 3 times greater than for design B, then we see that $N_{\rm ast,A} \psi_{A}^{-2/3} N_{\rm total,A}^{-1/3} = (2 N_{\rm ast,B}) (3 \psi_{B})^{-2/3} (2 N_{\rm total,A})^{-1/3}$, i.e. the probability of discovery for design A is only $76\%$ of the probability for design B, making design B favorable.

\subsection{TeV search for continuous point sources}
\label{section:tevcontinuous}

For TeV muon-neutrino searches, we have high detection rate of background neutrinos, making detection feasible only if $N_{\rm s} \gg 1$. This requires a different treatment than what we saw above for the ultra-high-energy search.

\subsubsection{Discovery probability}

The expected number of directional coincidences between detected neutrino events and the sources in the catalog, assuming no signal component and therefore only chance coincidence, is
\begin{equation}
\langle N_{\rm c} \rangle = k \psi^2 \rho_{\rm source} d_{\rm th}^3 N_{\rm total}
\label{eq:Nc}
\end{equation}
where $k$ is an appropriate numerical constant. For TeV searches, $N_{\rm total} \gg 1$ for typical observation periods; therefore the probability distribution of the measured number $N_{\rm c}$ of coincidences, still assuming no signal component, can be approximated with a normal distribution with $\langle N_{\rm c} \rangle$ mean and $(\langle N_{\rm c} \rangle)^{1/2}$ standard deviation. For normal distributions, $p_{\rm disc}$ can be expressed as a minimum offset $\sigma_{\rm disc}(\langle N_{\rm c} \rangle)^{1/2}$ from the mean. For instance, for $5\sigma$ discovery we need $\sigma_{\rm disc} = 5$. The probability of discovery is therefore equivalent to the probability that the number of coincidences will be $\geq \langle N_{\rm c} \rangle + \sigma_{\rm disc}(\langle N_{\rm c} \rangle)^{1/2}$.

We are interested in calculating the probability $P(p_{\rm disc})$ of reaching a p value $\leq p_{\rm disc}$ in the presence of $N_{\rm s}$ signal neutrinos. For simplicity, we can approximate $N_{\rm b} \approx N_{\rm total}$ since we will only use sources within the local universe, representing a small fraction of the astrophysical neutrino flux. This means that the expected number of coincidences will be shifted by $\approx N_{\rm s}$ in the presence of signal neutrinos. We therefore have an approximately normal distribution of coincidences with $\langle N_{\rm c} \rangle + N_{\rm s}$ mean and $(\langle N_{\rm c} \rangle)^{1/2}$ standard deviation. The probability of this distribution exceeding the discovery threshold, that is the probability of discovery, is
\begin{equation}
P(p_{\rm disc}) = 1 - \Phi\left(\frac{\sigma_{\rm disc}\sqrt{\langle N_{\rm c} \rangle} - N_{\rm s}}{\sqrt{\langle N_{\rm c} \rangle}}\right)
\label{eq:pdisc2}
\end{equation}
where $\Phi$ is the cumulative standard normal distribution function.

\subsubsection{Scaling relation}

While $P(p_{\rm disc})$ in Eq. \ref{eq:pdisc2} is a nontrivial function of the argument of $\Phi$, it is clear that it is a monotonically increasing function of $N_{\rm s} / (\langle N_{\rm c} \rangle)^{1/2}$ and only through this combination does it depend on the detector parameters. This combination gives us the scaling relation
\begin{equation}
P(p_{\rm disc})|_{\rm continuous,TeV} \longrightarrow N_{\rm ast} \psi^{-1} N_{\rm total}^{-1/2}.
\label{eq:ppdisccontinuoustev}
\end{equation}
Note that this is not proportionality, but a monotonically increasing dependence, hence the $\longrightarrow$ notation instead of $\propto$. Nevertheless, detector design optimization with respect to continuous TeV searches requires the maximization of the right hand side of Eq. \ref{eq:ppdisccontinuoustev}.

\section{Transient sources}
\label{section:transients}

Transient high-energy processes represent some of the most interesting cases for the production of high-energy neutrinos. Primary examples for transients that may be important neutrino sources are gamma-ray bursts (GRB), certain types of supernovae, magnetar flares and AGN flares. The transient nature of these sources greatly facilitates neutrino searches enabling the use of temporal coincidence, reducing the background rate.

Detector optimization for transients, as we discuss below, is very similar to the case of continuous sources. This is good news as it allows for optimization synchronously for the two source categories. Below we discuss the cases of TeV and ultra-high energy observations, as well as a third category, the case of rare events that requires a different optimization.

\subsection{Rare transient events}
\label{section:raretransient}

For continuous sources, we assumed that there is a sufficiently large number of known sources in the catalog such that, in order to reach the required false alarm probability $p_{\rm disc}$, we need to reduce the source catalog by excluding some of the sources. This may not always be necessary for rare transient sources.

We can quantify the requirement towards a source type for a single coincidence being a discovery. Assuming that the probability of temporal and directional coincidence is small, it is given by
\begin{equation}
p \approx k \psi^2 R_{\rm source} \tau_{\rm source} N_{\rm total}
\label{eq:ptransient}
\end{equation}
where $R_{\rm source}$ and $\tau_{\rm source}$ are the overall rate and the typical duration of the source, respectively, and $k$ is an appropriate numerical constant.

In Eq. \ref{eq:ptransient} all the observed transients are included in $R_{\rm source}$. This means that, if $p$ is less than the required false alarm probability $p_{\rm disc}$ for discovery, then we can keep all sources in the analysis. Consequently, the only factor that will affect our chance of discovery is the sensitivity of the detector, i.e. the number of signal neutrinos detected. This gives us the scaling relation
\begin{equation}
P(p_{\rm disc})|_{\rm transient,rare} \propto N_{\rm ast}.
\end{equation}

For which scenarios can we consider the transient sources sufficiently rare to assume the above scaling? Let us consider the different cases.

For GRBs, we have a rate of $R_{\rm grb} = O(100)$,/yr$^{-1}$ and duration $\tau_{\rm grb} \sim 100$\,s. For the case of ultra-high energy neutrinos, including cascade events we can expect $N_{\rm total} \sim 200$ detections, assuming a $10$-fold increase compared to the current IceCube rate, within a one-year period. For $\psi \sim 15^\circ$, using Eq. \ref{eq:ptransient} we get $p \approx 0.004 \ll 1$, therefore GRBs can be safely considered rare for ultra-high-energy searches. For the case of TeV searches, assuming $\psi \sim 1^\circ$ and $N_{\rm total} \sim 10^6$ representing a 10-fold increase compared to current IceCube rates, we get $p = 0.1 \ll 1$. GRBs are therefore rare events from the perspective of TeV searches as well.

Examining other transient sources in a similar fashion, we find that magnetar flares can be considered rate events. While they are more numerous than GRBs, we can only detect nearby events, making their overall rate lower, while their duration is comparable to those of GRBs. Supernovae and AGN flares, however, are more numerous and last for significantly longer, making them sufficiently common to require a more detailed analyses. The only exception is UHE searches with only track events. For this scenario, practically all transients can be considered rare.

\subsection{Ultra-high-energy and TeV searches for transient sources}
\label{section:transients}

For transient sources that are not sufficiently rare to merit the above simple treatment, we find scaling relations building on the derivations for continuous sources.

For this, we point out that Eq. \ref{eq:ptransient} is equivalent to Eq. \ref{eq:p} for continuous sources with $R_{\rm source} \tau_{\rm source}$ in the transient case corresponding to the number of sources $N_{\rm source} = 4/3 \pi \rho_{\rm source} d_{\rm th}^3$ in the continuous case. Similarly, Eq. \ref{eq:Nc} can be also expressed for transients with the same substitution. Besides this difference, the derivations of the scaling relations for the ultra-high-energy and the TeV cases can be carried out identically for the transient and continuous cases. Consequently, the scaling relations for the two cases will also be identical:
\begin{eqnarray}
P(p_{\rm disc})|_{\rm transient,UHE} \propto N_{\rm ast} \psi^{-2/3} N_{\rm total}^{-1/3} \\
P(p_{\rm disc})|_{\rm transient,TeV} \longrightarrow N_{\rm ast} \psi^{-1} N_{\rm total}^{-1/2}.
\end{eqnarray}

\section{Extended galactic source}
\label{section:galactic}

Another interesting scenario to consider is the contribution of galactic sources to the astrophysical neutrino flux. While galactic sources cannot explain the total extraterrestrial neutrino flux detected by IceCube, they an still contribute to it. Since galactic sources would be much easier to identify and study then many extragalactic sources, even a partial contribution to the overall flux would be very interesting.

Here we consider an extended (unresolved) galactic region as the potential origin of some astrophysical neutrino flux to derive scaling relations for the detector parameters that optimize the detectability of such neutrino flux. Such extended source can be, for instance, the vicinity of the galactic center.

For simplicity, we assume that the extended region is circular (e.g., galactic center) with angular radius $\psi_0$. Treating this radius as an angular uncertainty from the center of the extended region, we have an effective angular uncertainty of $\psi_0 + \psi$ for the neutrinos. Beyond this, the scaling of $P(p_{\rm disc})$ for this case can be derived similarly to the case of continuous point sources. Given that we expect the galaxy to represent only a small fraction of the detected astrophysical flux, we are likely in the regime in which multiple detected neutrinos are required to claim discovery from the extended region. This means that the scaling will be similar to that of the continuous TeV case, independently of the considered neutrino energy range:
\begin{equation}
P(p_{\rm disc})|_{\rm galactic} \longrightarrow N_{\rm ast} (\psi_0 + \psi)^{-1} N_{\rm total}^{-1/2}.
\label{eq:ppdisccontinuoustev}
\end{equation}

\section{Conclusion}
\label{section:summary}

We derived source model independent scaling relations of detection probability for high-energy neutrino detectors with respect to detector parameters. The considered parameters we considered are (i) the number $N_{\rm ast}$ of detected astrophysical neutrinos from a given source population, (ii) the angular uncertainty $\Psi$ of reconstructed neutrino directions, and (iii) the total number $N_{\rm total}$ of detected neutrinos, including those of atmospheric as well as astrophysical origin. Table \ref{table1} summarizes the obtained scaling relations for easier comparison.

\begin{table}
\begin{center}
\bigskip
\bigskip
\begin{tabular}{| c ? c | c | c | c | c |}
\hline
Source type & Search & $N_{\rm ast}^{(\cdot)}$ & $\Psi^{(\cdot)}$ & $N_{\rm total}^{(\cdot)}$ & Section\\ \hlinewd{1pt}
Continuous  & UHE    &       1             &     -2/3     &    -1/3      &   \ref{section:UHEcontinuous}      \\
\cline{2-6} & TeV    &       1             &     -1       &    -1/2      &   \ref{section:tevcontinuous}      \\ \hline
Transient   & rare   &       1             &      0       &     0        &   \ref{section:raretransient}      \\
\cline{2-6} & UHE    &       1             &     -2/3     &    -1/3      &   \ref{section:transients}         \\
\cline{2-6} & TeV    &       1             &     -1       &    -1/2      &   \ref{section:transients}         \\ \hline
Galactic    & all    &       1             &     $*$      &    -1/2      &   \ref{section:galactic}           \\ \hline
\end{tabular}
\caption{Scaling relations between detector characteristics $N_{\rm ast}$, $\Psi$ and $N_{\rm total}$ for different source types and neutrino searches. $^*(\Psi_0 + \Psi)^{-1}$.}
\label{table1}
\end{center}
\end{table}

Our conclusions are as follows. (i) The above scaling relations allow for a source \emph{model-independent} comparison of different detector design. (ii) Most scenarios result in relatively similar scaling relations, enabling optimization that only weakly depends on the target signal type. (iii) For the TeV cases, we recover the familiar signal-to-noise-ratio scaling $N_{\rm ast} / (\Psi^{2} N_{\rm total})^{1/2}$. For the UHE and rare searches, however, $N_{\rm ast}$ is more important. We can therefore conclude that, overall, detector sensitivity is particularly important in optimizing detector design.

\vspace{4 mm}
The author is thankful for Marek Kowalski and the IceCube Gen2 HEA/Surface working group for the useful feedback. We are thankful for the generous support of Columbia University in the City of New York.

\bibliographystyle{elsarticle-num-names}

\end{document}